\documentclass[useAMS,usenatbib,usegraphicx]{mn2e}
\title[ The historical light curve of AG Dra]
{The historical light curve of the symbiotic star AG Draconis:  intense, magnetically induced cyclic activity}

\author[Liliana Formiggini and Elia M. Leibowitz ]
{Liliana Formiggini$^{1}$\thanks{E-mail:
lili@wise.tau.ac.il}
and Elia M. Leibowitz$^{1}$\thanks{E-mail: elia@wise.tau.ac.il}\\
 $^{1}$The Wise Observatory and the School of Physics and Astronomy, Raymond
and Beverly Sackler Faculty of Exact Sciences \\ Tel Aviv University, Tel
Aviv 69978, Israel\\}

\def\simi{$\sim$}
\def\apm {$\pm$}

\def\Pa{{\em P}$_{a}$}
\def\Pb{{\em P}$_{b}$}
\def\Pr{{\em P}$_{r}$}
\def\Pm{{\em P}$_{m}$}
\def\Pp{{\em P}$_{p}$}

\begin{document}

\pagerange{\pageref{firstpage}--\pageref{lastpage}} \pubyear{2011}

\maketitle
\label{firstpage}

\begin{abstract}

We analyse the historical optical light curve of the symbiotic system AG Draconis, 
covering the last 120 years. During the first 31 years the brightness of the star 
has not been varying by more than 0.1 mag. A weak periodic signal with the binary 
period of the system of \simi 550 d can be detected in this section of the light 
curve, as well as in all other later quiescence sections of it.

Around the year 1922 the quiescence brightness of the star increased by 0.29 mag. 
Since then the star's photometric history is marked by a series of brightness 
fluctuations with an amplitude of 1-2 mag and a typical duration of 
100-200 d. The time intervals between outbursts are integral numbers of the period 
373.5 d. The outbursts are grouped in 6 dense clusters, each one lasting some 1500 d, 
that are well separated from one another along the time axis 
with a quasi periodicity of 5300 d. 

We suggest that the outbursts of the system are triggered by episodes of intense mass 
outflow from the atmosphere of the cool star onto the environment of the hot component. 
The 373.5 d cycle is the length of a "day" on the surface of the giant that rotates in 
retrograde direction with a sidereal period of 1160 d. A weak signal with this 
periodicity is also present in the light curve.  The modulation of mass transfer in 
the system is a combined effect of a dipole magnetic field of the giant star and the 
tides induced in its atmosphere by its binary companion. The 5300 d quasi period is 
that of a solar-like magnetic dynamo process that operates in the outer layers of 
the giant. The combined effect of the 5300 d and 373.5 d cycles induces a 
second mode of pulsation of the giant star with the period of 350 d. 

AG Dra is the 5th symbiotic system that shows in its historical light curve this 
type of intense magnetic and magnetically induced activity.

\end{abstract}

\begin{keywords} binaries: symbiotic -- stars: individual: AG Dra -- stars:
magnetic fields -- stars: rotation.
\end{keywords}

\section{ Introduction}
The star AG Draconis is one of the most intensively studied symbiotic systems. 
Nonetheless, its long term light curve (LC), in quiescence as well as during 
its outburst episodes, is far from being well understood.

Photometric data of AG Dra, recorded since the year 1890 display several active 
phases, separated by quiescence periods. In quiescence, a small amplitude 
modulation is detectable in the B and V band. In the U-band the variation 
amplitude  is larger, \simi 1 mag, and shows a  periodicity of 554 d 
(Meinunger 1979). Three values are given in the literature  for the orbital 
period of the system, as derived from radial velocity data: 554.0  
(Mikolajewska et al. 1995), 548.6   (Fekel et al. 2000b), 550.5 (Friedjung et al. 2003).

Many optical eruptions occurred since the first one recorded in the year
\simi 1930. Iijima (1987) suggested a 15 years periodicity for outburst 
occurrence.

\begin{figure*}
\includegraphics[width=120mm]{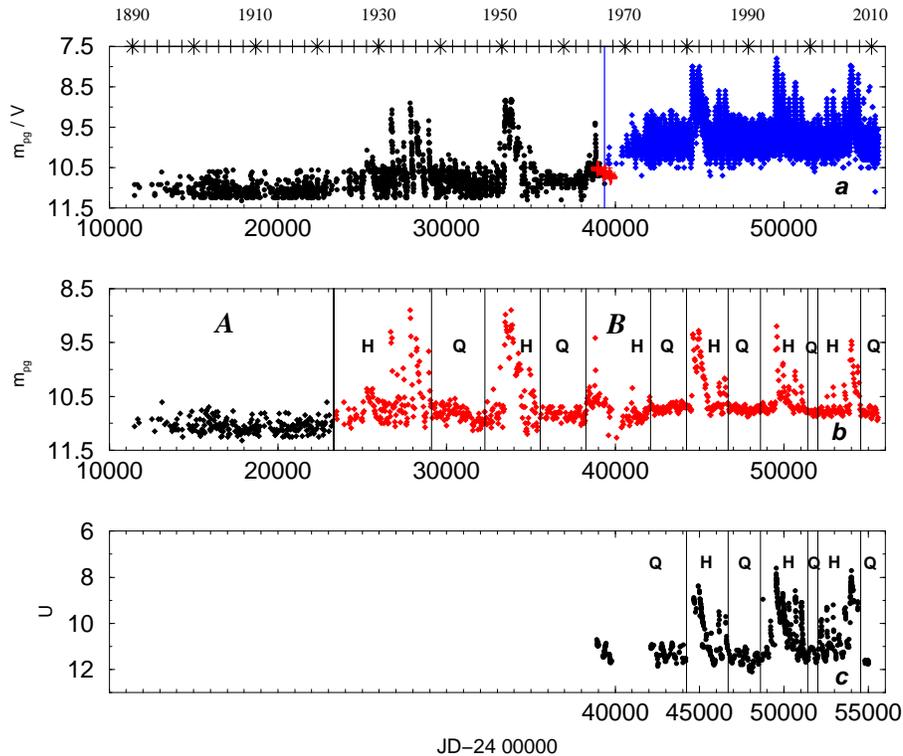}
\caption{The LC of  AG Dra, from the year 1890 up until Feb 2011. (a) Raw data of the 2 main sources, Harvard plates {\em m}$_{pg}$ to the left of the vertical line and AAVSO visual data to its right. The crosses mark the few Belyakina (1969) data. (b) The LC scaled to a common color system. The data are binned into 30 d equal width bins. A heavy vertical line marks the border between sections A and B of the LC. Thin vertical lines separate between Quiescence and High states of the system (see text for further explanations). (c) Photoelectric measurements in the U bandpass filter.}
\end{figure*}

The presence of more than one periodicity in the LC of AG Dra 
is claimed by several authors. Bastian (1998) analysing the data 
base of the Association Francaise des Observateurs d'Etoiles 
Variables (AFOEV) covering the years from 1973 to 1997, claimed 
the presence of a \simi 380 d period. This periodicity is
also detected in the radial-velocity data compiled 
by Mikolajewska et al. (1995).
Petr\'{i}k et al. (1998)  found {\em P}=552 d during the quiet state, and
{\em P}=350 d during the activity state. Friedjung et al. (1998)
detected three periods,  592 d in U, 356 d in B, and 350 d in V and 
suggested that the 350 d period could represent the cool giant's pulsation.
G\'{a}lis et al. (1999) found two periodicities, 549.73 d and 355.27 d
in photoelectric and radial velocity data.

In a more recent paper, Friedjung et al. (2003) reaffirmed the presence of a 
signal with a 357 d period in the radial velocity curve of the star, and
suggested again that the origin of this signal is in pulsations 
of the giant star of the system.

Motivated by the discovery of multiperiodicities in the LC of several symbiotic 
systems, BF Cyg (Leibowitz \& Formiggini 2006), YY Her (Formiggini \& 
Leibowitz 2006), Z And (Leibowitz \& Formiggini 2008) and 
BX Mon (Leibowitz \& Formiggini 2011) we have analysed the photometric behavior 
of AG Dra, with the aim of examining the overall periodic content of the LC 
of this system. 

In Section 2 we present the data sets used, and the method used for constructing 
a consistent LC along the 120 years of observations. 
Section 3 describes the time-series analysis and the periods detected. 
Section 4 discusses the proposed interpretation of the identified periods.  
In the summary we compare the results presented in this work with our past 
findings in the light curves of another 4 symbiotic systems, Z And, BF Cyg, 
YY Her and  BX Mon.

\section{The data}

The long-term LC of AG Dra is covered by two major data sources. One is
the photographic 
estimates from the Harvard patrol plates, obtained between 1890 and 1966 
(Robinson 1964, 1969). 
In our analysis, we used the data of Table 6, 7 and 8 of Robinson (1969),
spanning the interval from JD=24 14787.644 to JD=24 39355.398.  
To this we added a few data in the B band from Belyakina (1969).These 
measurements were transformed to {\em m}$_{pg}$ using the few points in the 
overlapping portion with the LC of Robinson (1969). 

The second large data set is the visual magnitude estimates collected by the 
American Association of Variable Stars Observers (AAVSO), covering the time 
interval from JD=24 39584 to JD=24 55600.

Fig. 1 a) shows the raw data of our two major data sources, one on the left and 
one on the right of the vertical line in the figure. 
The conspicuous large magnitude difference between them is obviously due to the 
different photometric systems of the two data sets.

A few other data sets, mostly of photoelectric photometry measurements have 
also been used by us, mainly for calibrating and establishing a common 
magnitude scale for the two major sets. This process is described in the 
next subsection. Note, however, that the major results of this work hardly 
depend on the calibration accuracy. 

\subsection{Scaling the data}

In order to create one consistent LC we had
to find a reliable transformation between the photographic system of the 
Harvard patrol plates and the eye-visual estimates of the AAVSO data. 
Unfortunately, there is no overlap in time between these two sets of data. 
Skopal (2007) estimated that the emission lines contribution to the star 
brightness in the V band amounts  to  \simi 0.12 mag. The variability of 
emission lines in the AG Dra spectrum as well as the beginning of an 
activity phase of the system (see below), make the task of scaling more 
complicated.

Consequently, instead of applying the numerical relation 
between  the earlier photographic magnitudes and the B and V magnitudes 
we adopted an empirical approach, taking advantage 
of the data collected by many campaigns of photoelectric observations of AG Dra, 
using them as a bridge between our two major sets. We retrieved all the  V and 
B photoelectric data obtained from the year 1980 up to 2011 
(Burchi 1980; Meinunger 1981; Iijima 1987; Luthardt 1992; Montagni et al. 1996;
Greiner et al. 1997; G\'{a}lis et al. 1999; Leedj\"{a}rv et al. 2004; Hric et al. 1991; 
Skopal et al. 1992; Hric et al. 1993, 1994ie; Skopal et al. 1995;
Hric et al. 1996; Skopal 1998; Skopal et al. 2002, 2004, 2007; Munari et al. 2009).

Four segments of quiescence state, which are contemporaneous with AAVSO data points, 
have been selected in the LC of the photoelectric data, namely 24 40680$<$JD$<$24 44200, 
24 45613$<$JD$<$24 49105, 24 51142$<$JD$<$24 52105, 24 54555$<$JD$<$24 54892.
We evaluated the empirical scale shift between the V photoelectric data 
and the eye-visual of the AAVSO data, calculating the weighted averages of the data 
in the four quiescence segments. Using the average (B-V) at quiescence, we transformed 
the V-scaled AAVSO data to B photoelectric magnitudes. Using the few simultaneous B 
and {\em m}$_{pg}$  data of Greiner et al. (1997), an empirical shift has been established
for scaling the B data to the old Harvard photographic system.

With this procedure  we transformed the AAVSO set to the scale of the {\em m}$_{pg}$ data.
We have also calculated anew the empirical shift factor for the Belyakina (1969) data.

Fig. 1 b) shows the final resulting LC  transformed to {\em m}$_{pg}$ passband. In this 
presentation we binned the  data points into  bins of 30 days width. 
We note, however, that the results of our following analysis are insensitive to any 
particular reasonable binning, including no binning at all, of this data set.

\subsection{The two sections of the Light Curve}

Two distinct sections are clearly identified in the LC seen in Fig. 1b. Section A, to 
the left of the heavy vertical line at JD 24 23300, consists all the measurements performed 
earlier than that time. Section B are measurements performed later. No outburst has 
been recorded during the 11800 days covered by Section A.

Section B is characterized by a sequence of major outbursts of the system, of an 
amplitude of 1 to 2 mag with a typical duration of some 100 days. These are grouped 
within 6 distinct clusters, well separated along the 32280  day duration of this section. 
The thin vertical lines in Fig. 1 b) mark the borders between these clusters, to which 
we refer also as high (H) states of the system, and the in between time intervals, 
to which we refer as the quiescence (Q) states. The results of our following analysis are 
insensitive to any reasonable variation in the position of these border lines.

The median magnitude of the Q LC of Section B is brighter by 0.29 mag 
than the median magnitude of Section A. The  jump in the quiescence 
luminosity of the star around the year 1922  is not an artifact, being measured 
within the single consistent data set of the Harvard plates.

In general we note that all results presented in this paper are  independent of 
the scaling procedure presented in  sec. 2.1, because Section A consists entirely of data of 
the Harvard plates system and section B is mostly the AAVSO data set. We obtained  almost 
the same results as for section B if we analyse only the AAVSO data.

\begin{figure*}
\includegraphics[width=130mm]{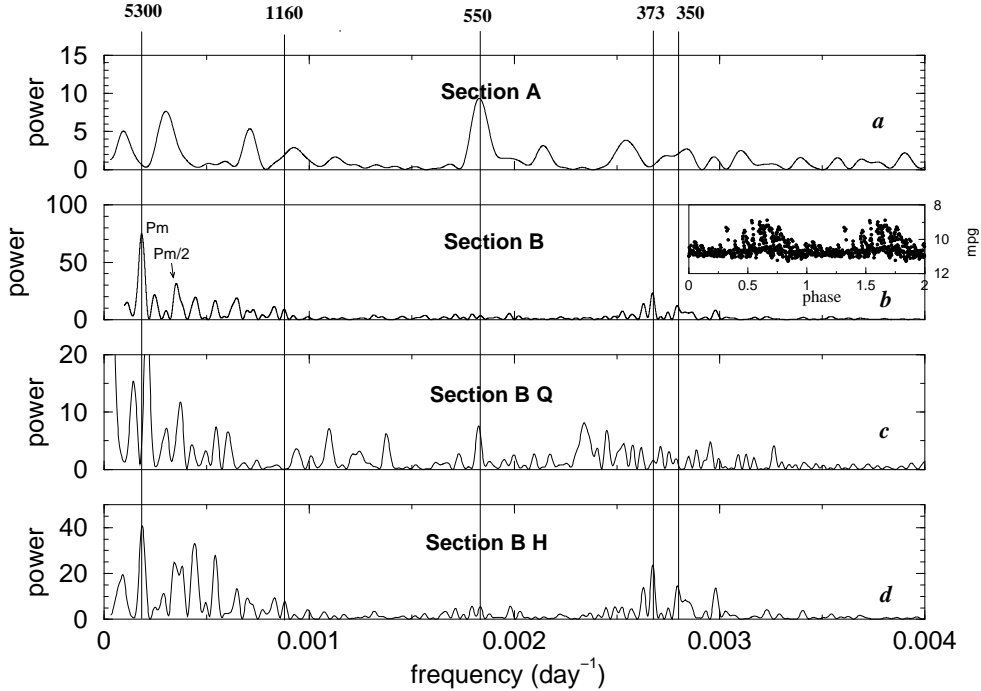}
\caption{(a) Power spectrum  (PS) of Section A of the LC.
 (b) PS of Section B. The inset shows the folding of Section B onto \Pm=5300 d. (c) PS of the quiescence segments of Section B. (d) PS of the outburst segments of Section B. Vertical lines show the 5 periods identified.}
\end{figure*}

\subsection{ The U light curve}

We have also analyzed a third data set of photoelectrical photometry measurements 
in the Johnson U band.
The sources are: Meinunger 1981; Taranova \& Yudin 1982; 
Iijima 1987; Martel \& Gravina 1985; Hric et al. 1991; Luthardt 1992; 
Skopal et al. 1992; Hric et al. 1993, 1994; Skopal et al. 1995; 
Montagni et al. 1996; Greiner et al. 1997; Skopal 1998; G\'{a}lis et al. 1999; 
Skopal et al. 2002, 2004; Leedj\"{a}rv et al. 2004; Skopal et al. 2007; 
Munari et al. 2009. 
Fig. 1 c)  presents the U LC of the system. It covers the last two clusters of 
outburst events 
seen in the photographic light curve shown in Figs 1 a) and b). The structure of the 
two events is very similar in the 2 photometric bands. Here again we refer to 
Quiescence (Q) states and High (H) states of the system as in Fig. 1 b).

\section{Time-Series Analysis}

\subsection{Section A and the quiescence states}

Section A of the LC covers a time interval of 11800 d during which the system was 
in a quiescence state, with no interruption by any outburst. The PS of this LC 
(Scargle 1982), shown in 
Fig. 2 a), has its highest peak at the period 546.5 d.  This peak is higher than 6 
times the standard deviation of the noise in the PS, but its false alarm 
probability (Scargle 1982) is not small. We consider it significant in view of 
the fact that a similar peak is present also  in the PS of the independent Q LC 
of section B, displayed in Fig. 2 c), at a frequency corresponding to 548 d. 
In the PS of the quiescence states of the U LC shown in Fig. 4 a), a similar peak, 
at a frequency corresponding to 547 d is present, where it has a clear high 
statistical significance (see section 3.3). Finally, the frequency of this peak 
coincides well within its uncertainty interval with the well known binary period
of the system. 

A second feature in the PS of section A is a broad peak around the period of 350 d.
Although by itself it is not statistically significant within this PS,
we draw attention to it, in view of its appearance in another independent LC, 
that of the PS of section B shown in Fig. 2 b), especially in the PS of the H 
states shown in Fig. 2 d). It is also clearly present in the PS of the H states 
in the U LC shown in Fig. 4 b) (see section 3.3). It also coincides with a 
periodicity that was identified in the radial velocity curve of the AG Dra 
system (Galis et al. 1999, Friedjung et al. 2003). The uncertainty in the value 
of this period is quite large, as apparent in the large width of the feature 
in the PS, due to interference with the 365 d periodicity in the timing of 
the observations. We refer to this period as \Pp= 350 d.

\subsection{Section B and the major six outbursts}

Fig. 2 b) displays the PS of section B of the LC in the period range 
250-30000 days. The major peak is around the frequency corresponding to 
the period 5300 d. Its false alarm probability (Scargle 1982) is less 
than 5 percent. The second highest peak corresponds to its 2nd harmonic. 
The inset shows twice the cycle of the B LC folded onto this periodicity. 
As apparent in the inset, as well as in Fig.s 1 a) and b), the \simi 5300 d
cycle is a quasi periodicity in the occurrence of the 6 clusters of 
outbursts along the last 88 years in the history of the star, referred to above.

The third highest peak in Fig. 2 b) corresponds to the period 373.4 d. 
The next highest peak to its left is an alias of it. It disappears from the PS 
when a "clean" routine is applied on the data. The 373 peak is even more 
pronounced in the PS of H states shown in Fig. 2 d). The broad feature to the 
right of the 373 peak corresponds to the \Pp=350 d periodicity mentioned above. 
    
\subsubsection{Statistical Tests and Simulation}

As a test of the statistical significance of the \Pm\simi 5300 d period, we 
have developed, following Broadbent (1955, 1956, and references therein), 
a period  search routine that finds the periodicity, or "quantum", in the 
language of Broadbent, that fits best  a set of given numbers. The time points 
among which we looked for a "quantum" are the highest points in the 6 clusters 
of the H state of the system. We performed this search on various different 
binning of the LC and the "quantum" found is \simi5300 d, with a dispersion of 
the observed high points around the predicted times for a strict periodicity 
is some 450 days, or 0.085 in the phase space.  

The 373 d periodicity in section B of the LC is clearly far from being harmonic. 
Its Fourier signal is therefore rather weak. We therefore applied on the same 
data the "quantum" search routine, which is especially suitable for finding a 
cyclic feature in a time series, regardless of the detailed structure of the 
time dependent parameter of the series.  Here the set of numbers for which a 
"quantum" is looked for  is the times of the highest points of the individual 
outbursts of the system that rise higher than $2.5\sigma $ of the magnitude 
population. This test enables also an evaluation of the statistical significance 
of the 373 d period. 

We find a best-fitted period ("quantum") of \Pa=373.5 d.  The dispersion is  
S=49.34 , which is 0.132 in phase. The time of the 1st peak is   JD=2425249.03.

The high statistical significance of the 373 d period is confirmed by  the value 
of the statistic $\sqrt(N)(1/3-S^{2})=1.87  > 1$. 
Here N=35 is the number of elements in the series of times of maximum brightness 
in individual outbursts of the star. As shown by Broadbent (1956) this inequality 
is indicating a very small false alarm probability for the 373 d period. 

As an additional check, we have also conducted a "bootstrap" (Monte-Carlo) test 
of the robustness of the 373 d periodicity (Efron and Tibshirani 1993).
On a sample of over 10000 pseudo-observed bootstrap sets we have not obtained 
even one case with a dispersion that is as small as in the real data. 
Thus, the probability of false alarm is smaller than 1/10000.

We performed the Broadbent period search technique and the accompanying test 
of statistical significance on the LC binned with various  binning systems. 
These include also binning into bins with varying bin width. The results 
were insensitive to the binning system.

Fig. 3 is a zoom plot of the B LC. The dot-dashed vertical lines 
mark the peaks of individual outbursts that we determine by a computer 
selection routine, as well as by eye inspection of the data points of the LC. 
The  vertical lines indicate the zero phase of the 373.5 d periodicity, 
nearest to each of the 35 selected peaks. 

Fig. 3 shows  that the \Pa=373.5 d periodicity is preserved throughout 
the 32800 days covered by the B LC.

\subsubsection{The \simi 350 d  and the 1160 d periods}

In Figs 2 b) and d) one can identify a small peak that corresponds to the  
period 358 d that is also seen in Fig. 2 a), the PS of section A. We 
identify it with the \Pa=350 d periodicity, already mentioned above.
This peak is also clearly seen in Fig. 4 b), the PS of the 
entirely independent data set of the U LC of the system (see section 3.3). 
The large variability in the exact peak position of this feature in the 
different PSa is due to interference with the nearby annual 365 d periodicity 
inherent in long term ground based observations.

Another small peak is marked in Fig. 2 b) and d), corresponding to the 
period \Pr=1160 d. It is not statistically significant within the PS of 
section B. However, its conspicuous appearance in the PS of the 
independent data set of the U LC (Fig. 4 b) implies with a high
statistical probability, that it does represent a true periodicity of the 
AG Dra system.

Application of a "clean" PS routine on the section A and section B data sets 
shows that within the signal to noise level of the data, it is rather 
unlikely that any additional periodicity is hidden in the data, in particular 
any periodicity that might explain the residual variability that still 
remains in the LC after removing from it all the periods listed in Table 1 below. 
This was also confirmed by numerical tests that we performed, showing that 
adding one or a few periods, beyond those listed in Table 1 does not change the 
quality of the curve fitting in any significant way.

\subsection{The photoelectric U light curve}

Fig. 1 c) presents the LC of the system as measured photoelectrically in 
the Johnson U band, extracted as described in section 2.3. It covers the 
last two clusters of outburst events seen in the visual light curve and 
shown in Figs 1a) 
and b), and having a very similar  structure. Here again we refer to 
quiescence (Q) and the outburst-high (H) states.
 
\begin{figure}
\includegraphics[width=90mm]{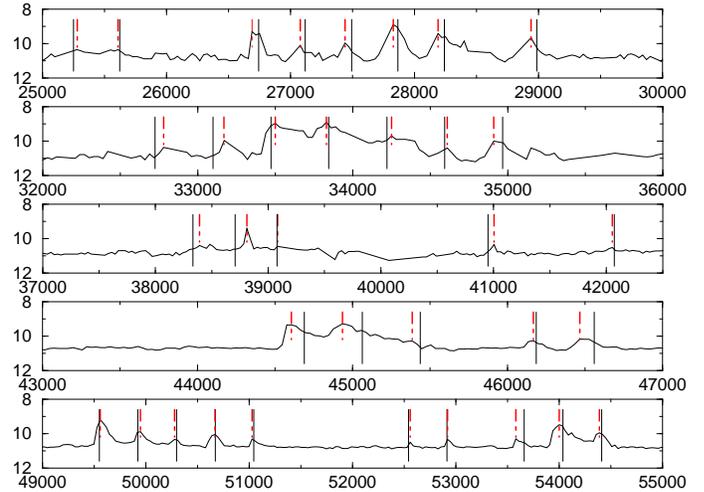}
\caption{ The 35 outburst peaks occurring in section B. The lines indicate the zero phase 
of a 373.5 d period, and the dot-dashed lines show the time of maximum light of each outburst.}
\end{figure}

Fig. 4 a) displays the PS of the U Q LC, and Fig. 4 b) that of its H section. 
The dominant peak corresponds to the period 547 d.  
The two satellites of the main peak are aliases of the 547 periodicity. 
The inset in Fig. 4 a) presents twice the Q  LC folded onto this periodicity. 
The other high peak in Fig. 4 b), corresponding to the period 1357 d, results 
from the interference of the 550 periodicity, the 5300 d periodic gaps in the 
partial Q LC, and the 365 d annual cycle: F(1357)=F(365)-[F(5300+F(550)]. 
Here F(P)=1/P.  

Fig. 4 b) is the PS of the U-H  LC. Here, in addition to the low frequency 
peaks due to the window function of the broken LC, the dominant peaks correspond 
to the periods 373.6 and 1130.4 d. The inset displays twice the H LC in the 
U filter, folded onto the 373.5 d periodicity.

\subsection{The periodic content of the LC}

Table 1 presents a summary of our findings concerning the periodic content in 
the long term LCs of AG Dra. There are 5 distinct periodicities that seem to 
dominate the temporal behavior of the continuum emission of the system. 
They are the quasi period 5300, and the periods 1160, 550, 373.5 and 350 days. 
The table presents our adopted value for each of these periods, and our 
estimated uncertainty in each one of them. Our error estimates are based mainly 
on the dispersions of the peak frequencies of the corresponding features in the 
different power spectra, and in the published literature. The last 5 columns in
the table indicates at which particular LC each period is present and the numerical 
value it takes in each of the corresponding PS. 

We note that all the variations that we are discussing in this paper can hardly 
be attributed to variation in emission lines. As noted already in section 2, 
Skopal (2007) estimated that the emission lines contribution to the star brightness 
in the V band amounts to  \simi 0.12 mag, while the main variations that are of 
our concern here are of the order of 1 mag. Furthermore, as Table 1 shows, all the 
periods discussed here appear to modulate the brightness of the system in 2 and 
even 3 different photometric broad passbands.
  
Table 1 presents also 2 numerical relations among the 5 periods of the system, 
which we shall discuss in the following sections.
 
\begin{table*}
\begin{tabular}{|llc||ccrcr|} \\
\multicolumn {8} {|c|}{\bf Table 1. Peaks in the  PSa}\\
\hline
 Name & Period (d) & Error  &\multicolumn {5}{|c|} {Measured period  in individual LCs } \\
\hline
\hline
     &        &            &  A & B-Q & B-H  & U-Q & U-H    \\
\hline
 \Pm &   5300 & quasi-p            &     &     & 5373  &      & 5310  \\
 \Pr &   1160 & \apm 40            &     &     & 1135  &      & 1130  \\
  \Pb &   550 & \apm 4             &546.5  & 548 &  545  & 547  &       \\
  \Pa=  &   373.5 & \apm 1           &     &   & 373.7 &      & 373.6 \\
 (1/\Pb+1/\Pr)$^{-1}$  &     &    &     &     &       &      &       \\
  \Pp=  &  350 & \apm 10           &352  &     &  358  &   & 344   \\
 (1/\Pm+1/\Pa)$^{-1}$ &   &      &     &     &       &      &       \\

\end{tabular} 
\end{table*}

\section{Discussion}

\subsection{The binary period}

Spectroscopic observations in AG Dra have established its binary orbital period 
between  548.5 d and 554.0  (Mikolajewska et al. 1995; Fekel et al. 2000b). 
This periodicity has been also clearly identified in 
photometric measurements in the U photometric band (Meinunger 1979).

As described in section 3.1 this photometric period  is present also at the two 
additional independent LCs of section A and of the quiescence states of section B. 
It is not detected during the high state events. It is also present in the U Q LC
(see Fig 4 a).

For period between 550.5 and 552.5 d, the phasing of the 3 independent 
photometric LCs presented in this work agree with each other to within 
less than 0.09 in phase space. 

We have also considered the radial velocity curve of the system, as presented by  
Mikolajewska et al. (1995), Fekel et al. (2000b).
The phase of maximum positive radial velocity for periods in the above period 
range is lagging by 0.25 to 0.3 in its value behind a weighted mean phase of 
minimum light of the corresponding 3 photometric LCs. Specifically, for P=552.5 
the phase difference is precisely 0.25. This is very much consistent with the 
"reflection" model that has been proposed as an interpretation of the U binary 
LC (Kenyon 1986, Formiggini \& Leibowitz 1990,  Skopal 2001).

\subsection{The other  periodicities}

As shown in Fig 2 b) and Table 1, section B LC is characterized by 4 periodicities. 
One is \Pm \simi5300 d, which is the characteristic quasi-periodicity 
in the occurrence of the 6 H states discussed in section 3.2. The other three, 
\Pa=373.5 d, \Pp=350 d and \Pr$\simeq$1160 d seem to  be the periods of three other cyclical 
processes that operate in the AG Dra stellar system, preserving their value as 
well as their phase throughout the time of the observations.

\subsubsection{The rotation period of the giant}

We suggest that the period \Pr$\simeq$1160 d is the sidereal 
rotation period of the cool component of the AG Dra binary system. 
This is different from estimation of the rotation period of the 
giant given  by  Zamanov et al. (2007). 
We note however that this estimate is based on a few assumptions and measured 
quantities that are far from being well established. They rely on the assumption 
that the axis of the giant rotation is normal to the binary plane, and they use an 
estimate of the radius of this star that is much in dispute in the 
literature (Mikolajewska et al. 1995, Zhu et al. 1999, Greiner et al. 1997, 
Huang, Friedjung \& Zhou 1994). 

Also, the uncertainty in the basic observational 
parameter of the projected equatorial radial velocity of the giant v${_r}$ sin i  
itself is quite large (Fekel, Hinkle \& Joyce 2004). According to 
Mikolajewska et al. (1995), strictly from observations, only an upper limit of 
the inclination angle of the orbital plane i(B)$<$70 $^{\circ}$  may be 
inferred. Also Kenyon \& Garcia (1986) analysing the parameters of T CrB system, 
caution that v${_r}$ sin i is not necessarily a true measure of the star rotation 
in such composite systems.

Some indirect evidence that 1160 d is the rotation period of the cool giant 
comes from the fact that it is absent from the LC of the
system in the quiescence states, in the visual as well as in the U
photometric bands (sections 3.1 and 3.3). We shall argue below, that the
quiescence of the system are time intervals during which the  magnetic
activity in the outer layers of the giant is dormant. At this times
the giant surface brightness is relatively uniform, with too few spots to 
enable identification of rotation by photometry. High states of the systems
are times of highly intensive magnetic activity, accompanied, as in
the sun, by appearance of dark spots on the  surface of the star. These are 
the features  that  enable photometric monitoring of the giant rotation.

Our suggestion that \Pr$\simeq$1160 d is the giant rotation period makes AG Dra 
an exception in the claimed general trend of synchronization of the star 
rotation with the binary orbital revolution, in symbiotic systems with 
orbital periods $<$1200 d, (Zahn 1977, Zamanov 2011). However, in the 
calculations of Zahn (1977) it is assumed specifically that the amplitude 
of the tidal oscillations of the  giant are small enough to 
justify linear treatment of the problem.  In AG Dra, the oscillations 
need not be small due to the thick stellar wind of this star (Skopal 2001).
Also the giant pulsations with the \Pp=350 d periodicity introduces a 
mechanical and hydro and thermo-dynamical elements into the physical 
processes that are so far unaccounted for by present day theory 
(see also Ogilvie \& Lesur 2012).

The rotation of the giant of AG Dra is not very unique in not being 
synchronized with the orbital revolution also on a purely observational basis. 
There are in fact some other symbiotic systems that 
are far from synchronization (see, for example Figure 1 in Zamanov 2011, 
Cikala et al. 2011). 

\subsubsection{The 350 d period}

As already noted in section 3.5, Friedjung et al. (2003) established a 
periodicity of \simi 350 d in the radial velocity of the system which they 
have attributed to pulsations of the giant star. In this work we discovered 
the same period in the LC of the system. Fekel et al. (2000b) have objected 
to this interpretation on the ground that the amplitude of the radial velocity 
oscillation with this period in AG Dra is much larger than those measured 
in pulsations of other single K giants with similar long periods. 

We shall return to this point at the end of section 4.3.

\subsubsection{The 373 d periodicity}

A major hint toward understanding the nature of the \Pa=373.5 d periodicity  can 
be found in the numerical relation that exists among this period and the other 
2 coherent periods of 
the system, namely: F$_{Pa}$=F$_{Pr}$+F$_{Pb}$ (Table 1). Here  F=1/P, 
is the frequency corresponding to the period P. This is an expression of the 
fact that the period \Pa ~in the 
occurrence of the major outbursts of the system is a beat of the binary 
period \Pb ~and the giant rotation period \Pr. 

A similar case of the presence of a beat of the binary and the giant rotation 
periods has been recently found by us in the LC of another symbiotic system 
BX Mon (Leibowitz \& Formiggini 2011). We propose that the interpretation that 
was given to the presence of the three periods in the LC of BX Mon is 
applicable also in the case of AG Dra.

If the giant star is rotating with the period \Pr ~in the retrograde sense 
with respect to the binary revolution, the period \Pa ~is the length of 
a "day" for an observer on its surface, whose sun is the hot 
component. Fekel, Hinkle \& Joyce  (2004) estimated that the radius of 
the Roche lobe of the giant is  141 R$\odot$. If the radius of the giant 
is indeed 70 R$\odot$ or more, as suggested by Greiner et al. (1997) and 
certainly if it is a supergiant, as suggested by Huang, Friedjung \& Zhou 
(1994), its surface layers are quite deformed by the tidal force 
exerted by its companion. In particular, its deformation can be described 
to first approximation as a bulge in its atmosphere that circulates around the 
surface of star with its synodic diurnal period, namely with the 
period \Pa (Lecar, Wheeler \& McKee 1976). 

\begin{figure}

\includegraphics[width=90mm]{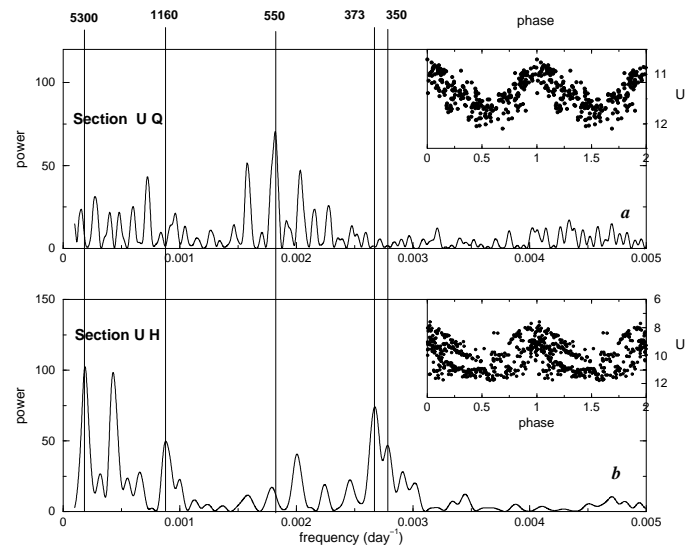}
\caption{(a)  PS of the quiescence U segments of the LC of AG Dra in the Johnson U band. The inset shows twice the Q U data folded onto the \Pb period. (b) PS of the outburst U segments. The inset shows twice the H U LC folded onto the 373 d periodicity. Vertical lines are as in Fig. 2.}
\end{figure}

\subsection{The proposed scenario}

We suggest that the giant star in the AG Dra system possesses a magnetic field, 
and that the axis of its dipole moment is inclined with respect to the rotation axis. 
Around the magnetic poles, due to the intensity and direction of the magnetic 
field lines, the hydrostatical equilibrium in the atmospheric layers is 
reached with material that is less bound by gravity than in areas where the 
magnetic field is weaker. Whenever the tidal bulge is crossing the area 
around one of the poles, equilibrium is broken and the giant is shading 
hydrogen rich material onto the Roche lobe of the hot component. A large 
amount of gravitational energy is being released in this  process, some of 
it is translated into the form of outbursts in visible light. The amount of 
material that is being poured onto the hot component in each one of such 
an accretion event would depend on the intensity of the magnetic field at 
the time of the bulge sweeping, as well as on hydro and thermo-dynamical 
parameters that characterize the bulge in the giant atmosphere at that time. 
In particular, it will depend, for example, on the phase of the giant peculiar 
pulsation period \Pp ~at the time of the mass transfer.  In events of large mass 
transfer,  hydrogen rich material may also reach the surface of the white dwarf 
of the system, igniting or intensifying nuclear reaction on its surface. 
Such events may be responsible for the particularly large outbursts of the 
system. 

This process has been termed "combination nova" by Sokoloski et al. (2006), 
who studied it in details in their thorough analysis of photometric and a 
great deal of spectroscopic data, covering two outbursts of the symbiotic 
star Z And. Recently a great deal of spectroscopic data has been accumulated 
about AG Dra, in the optical as well in in the UV and the x-ray regions of 
the spectrum (e.g. Shore et al. 2010). These provide a promising raw material 
for similar studies in AG Dra which may either confirm or perhaps refute 
the qualitative model that we are suggesting in this paper.

\begin{table*}

\begin{tabular}{@{}lrrrcrcrcrc@{}} \\
\multicolumn {11} {|c|}{\bf Table 2}{Comparison of the properties of Z  And, BF Cyg,
YY Her, BX Mon and  AG Dra}\\

\hline
  &  Z And & Ref.  & BF Cyg &Ref.  & YY Her & Ref. & BX Mon & Ref.& AG Dra & Ref.  \\

\hline
\\
Giant Sp. Type          &  M4 III      & 1    &  M5 III      & 1    & M4 III   & 1 & M5 III   & 1 &   K2 II   & 3\\
Binary period  (d)         &   759.0      & 5    &  757.3       & 6    &   593.2  & 7 &   1256   & 4 &   548.65   & 5\\
Giant Spin period (d)      &   482$^{*}$  & 8    &  798.8       & 6    &   551.4  & 7 &   656    & 9 &   1160   & 10\\
Tidal wave period (d)      &   1317       & 8    & 14580        & 6    &   7825   & 7 &   1373   & 9 &   373.5   & 10\\
Solar-type period (d)      &   7550       & 8    &  5375        & 6    &  4650    & 7 &   7370   & 9 &   5300   & 10\\

\\
\hline
\end{tabular}\\

*) This is a slightly preferred number among four possible values.

1) M\H{u}rset \& Schmid (1999), 2) M\H{u}rset et al. (1991), 3) Zhu et al. (1999), 4) Fekel et al. (2000a), 5)  Fekel et al. (2000b), \\
6) Leibowitz \& Formiggini (2006), 7) Formiggini \& Leibowitz (2006), 8) Leibowitz \& Formiggini (2008), \\
9) Leibowitz \& Formiggini (2011), 10) This paper. \\

\end{table*} 

As an example, we examined the peculiar phenomenon discovered by
Shore et al. (2010) that was termed by them the "bifurcation" of the O VI
Raman line in the bottom right frame of their figure 2 . We found 
that  the  six points that are responsible for this
effect, are measurements that were performed within less than .1 in phase 
according to the ephemeris of \Pa=373.5 (see section 3.2.1). 
In fact, Shore at al. already pointed out 
that these points are associated with an outburst of the system.
In view of our proposed cycle it will be interesting to check out whether
or not this effect appears cyclically in the spectrum with the period 
of 373.5 d.

Also the classification of the outbursts of AG Dra into hot and cool
types (Gonz\'{a}lez-Riestra et al. 1999) may  be related to the
difference that Sokoloski et al. (2006) have found between  
"combination nova" and disk instability outbursts in the LC of Z And.

According to our suggested scenario, the quasi-periodicity of \simi5300 d in the 
occurrence of the outburst events, is the period of a solar-like magnetic cycle 
that operates in the outer layers of the giant star. 
The quasi-periodicity of this cycle around 5300 is not unlike the 
well known quasi-periodicity of the solar magnetic cycle which has the mean period 
of 4000 d (11 years) but with individual cycles that vary between 3000 and 5000 
d (Babcock 1961; Mursula \& Ulich 1998; Fligge, Solanki \& Beer 1999).

The mass transfer episodes from the giant with the {\em P}$_{a}$ periodicity are 
most intensive when the intensity of the dipole magnetic field around the pole 
of the giant is highest. 
Hence the 5300 d quasi-periodicity in the  occurrence of the 6 clusters of outbursts in 
the history of the star.

The 373.5 d periodicity is a product of the parameters, 
the masses of the components, the radius of the giant, the inter-binary 
distance and the angle between the rotation and the magnetic axes of the giant. 
All these are independent of the internal magnetic dynamo process in the giant, 
hence the independence of the 373.5 d periodicity from the 5300 d one. 

According to this scenario, the brightening of the system by 
0.29  mag around JD 24 23300 is also related to the commencement of the operation 
of the magnetic dynamo solar-like cycle in the outer layers of the giant, through 
some process that is yet unknown.

Mikolajewska et al. (1995) also suggested that the outbursts of AG Dra are powered 
mainly by nuclear runway on the surface of the white dwarf (WD) of the system. 
These authors, however, were concerned about the fact that a continuous process 
of accretion of matter from the giant wind onto the WD may not 
be enough to fuel the eruptions of the system at the rate at which they are observed. 

Our suggested scenario is very much in line with Mikolajewska et al. idea. 
It adds a natural explanation for the cyclic nature of the outbursts. It also 
reveals the power source that is required in order to maintain the long term 
energy budget of the proposed process. Wind accretion from the giant is 
indeed not enough. The outbursts are fueled mostly by the flow of large 
amount of giant material through the L1 point of the system, at epochs of 
intense magnetic activity in the giant's outer layers.

Finally we note that the exceptionally large radial velocity amplitude of 
the pulsation of the K giant of AG Dra (Fekel et al. 2000b) may be well 
understood in view of the relation F(373)+F(5300)=F(350) (see Table 1). 
What gives the oscillations of the K giant of AG Dra their exceptionally 
large amplitude is the combined effect of the 5300 quasi periodicity of the
magnetic dynamo operating within the giant, and the 373.5 period of its tidal 
oscillations. These two cycles beat with each other, augmenting in the giant 
its pulsations mode of the 350 d periodicity.

\subsection {Comparison between BX Mon and AG Dra}

It is interesting to compare our findings in the LC of AG Dra with those 
found by us in the LC of BX Mon (Leibowitz \& Formiggini 2011). That star 
exhibits in its long term LC a series of outbursts that take place with a 
periodicity of 1373 d which, as in AG Dra, is the beat of the binary and 
the giant rotation periods of that system. As in section A of the LC of 
AG Dra, in BX Mon there is also a long time interval during which the 
1373 d periodicity in the occurrence of outbursts disappeared from the LC. 
But unlike the case in AG Dra, in BX Mon at that time, the 1373 period 
is being replaced by P=1263 d, the binary period of the system. 
This difference between the two symbiotics finds a natural explanation 
within the framework of the scenario that we are suggesting.

The disappearance of the spin-orbit beat period from the LC in both cases 
could be due to a long minimum phase in the operation of the magnetic dynamo 
cycle within the giant star, reminiscence of the well known Maunder minimum 
(Eddy 1976) in the history of the solar magnetic dynamo. 
At that minimum the dipole magnetic field of the giant is too weak to 
control the mass loss rate from the giant. In BX Mon, where the eccentricity 
of the binary system is \simi0.5 (Fekel et al. 2000a), the cyclically 
varying distance between the two components of the system 
takes over in modulating the mass transfer rate in the system with the 
binary periodicity. In AG Dra the eccentricity of the system 
is \simi 0 ~(Fekel et al. 2000b). Therefore, in the absence of the 
magnetic controlling process there is no other modulation mechanism for 
the mass loss from the giant, hence the lack of any outburst in 
section A of the LC of this star.

\section{Summary and the Generality of the phenomenon}

AG Dra is the 5th symbiotic system in the historical LC of which we have 
discovered similar patterns of temporal behavior. Here and in the other 
cases we interpret the findings as tracks of strong periodic activity, 
driven and modulated by three clocks in the system, the binary revolution, 
the giant rotation and a quasi-periodic, solar-like magnetic dynamo cycle 
in the outer layers of the giant star.  Table 2 presents the  
periodicities and quasi-periods that we have uncovered in the LCs of 
these 5 symbiotics. The similarities in the corresponding periods seem 
to be quite remarkable.

\section*{Acknowledgments}

       We acknowledge with thanks the variable star observations from the AAVSO 
International Database contributed by observers worldwide and used in this
research. We thank the referee, Dr. Steven Shore, for some very helpful comments.

\section*{References}

\def\ref{\par\noindent\hangindent 20pt}

\ref Babcock  H.W., 1961, ApJ, 133, 572
\ref Bastian U., 1998, A\&A, 329, L61
\ref Belyakina T.S., 1969, Izv. Krym. Astr. Obs., 40, 39
\ref Broadbent S.R., 1955, Biometrica, 42, 45 
\ref Broadbent S.R., 1956, Biometrica, 43, 32 
\ref Burchi R., 1980, IBVS, 1813, 1
\ref Cikala M., Mikolajewski M., Tomov T., Kolev D., Georgiev L., Munari U., Marrese P., Zwitter T., 2011, arXiv: 1102.5211v1
\ref Eddy J.A., 1976, Science, 192, 1189
\ref Efron B., Tibshirani R. J., 1993, An Introduction to the Bootstrap, Chapman \& Hall, New York 
\ref Fekel F.C., Joyce R.R., Hinkle K.H., Skrutskie M.F., 2000a, AJ, 119, 1375
\ref Fekel F.C., Hinkle K.H., Joyce R.R., Skrutskie M.F., 2000b, AJ, 120, 325
\ref Fekel F.C., Hinkle K.H., Joyce R.R., 2004, in Maeder A., Eenens, P., eds, Proc. IAU Symp. 215, Stellar Rotation. Astron. Soc. Pac., San Francisco, p.168
\ref Fligge M., Solanki S.K., Beer J., 1999, A\&A, 346, 313
\ref Formiggini L., Leibowitz E.M., 1990, A\&A, 227, 12
\ref Formiggini L., Leibowitz E.M., 2006, MNRAS, 372, 1325
\ref Friedjung M., Hric L., Petr\'{i}k K., G\'{a}lis R., 1998, A\&A, 335, 545
\ref Friedjung M., G'alis R., Hric L., Petr\'{i}k K., 2003, A\&A, 400, 595
\ref G\'{a}lis R., Hric L., Friedjung M., Petr\'{i}k K.,  1999, A\&A, 348, 533
\ref Garcia M.R., 1986, AJ, 91, 1400
\ref Greiner J., Bickert K., Luthardt R., Viotti R., Altamore A., Gonz\'{a}lez-Riestra R., Stencel R.E., 1997, A\&A, 322, 576
\ref Gonz\'{a}lez-Riestra, Viotti R.F, Iijima T., Greiner J., 1999, A\&A, 347, 478
\ref Hric L., Skopal A., Urban Z., Dapergolas A., Han\v{z}l D., Isles J.E., Niarchos P., 1991, Contrib. Astron. Obs. Skalnate Pleso, 21, 303
\ref Hric L. et al., 1994, Contrib. Astron. Obs. Skalnate Pleso, 24, 31
\ref Hric L. et al., 1996, Contrib. Astron. Obs. Skalnate Pleso, 26, 46
\ref Hric V. et al., 1993, Contrib. Astron. Obs. Skalnate Pleso, 23, 73
\ref Huang C.C., Friedjung M., Zhou Z.X., 1994, A\&A Supp., 106, 413
\ref Iijima T., 1987, Ap\&SS, 131, 759
\ref Kenyon S.J., 1986, The symbiotic stars, Cambridge Univ. Press, Cambridge, p.27
\ref Kenyon S.J., Garcia M.R., 1986, ApJ, 91, 125
\ref Lecar M., Wheeler J.C., McKee C.F., 1976, ApJ, 205, 556
\ref Leedj\"{a}rv L., Burmeister M., Mikolajewski M., Puss A., Annuk K., Galan C., 2004, A\&A, 415, 273
\ref Leibowitz E.M., Formiggini L., 2006, MNRAS, 366, 675
\ref Leibowitz E.M., Formiggini L., 2008, MNRAS, 385, 445
\ref Leibowitz E.M., Formiggini L.,  2011, MNRAS, 414, 2406
\ref Luthardt R., 1992, Mitt. Ver. Sterne, 12, 122
\ref Martel M.T., Gravina R., 1985, IBVS, 2750, 1
\ref Meinunger L., 1979, IBVS, 1611, 1
\ref Meinunger L., 1981, Mitt. Ver. Sterne, 9, 67
\ref Mikolajewska J., Kenyon S.J., Mikolajewski M., Garcia M.R., Polidan R.S., 1995, AJ, 109, 1289
\ref Montagni F., Maesano M., Viotti R., Altamore A., Tomova M., Tomov N., 1996, IBVS, 4336, 1
\ref Munari U. et al., 2009, PASP, 121, 1070
\ref M\"{u}rset U., Nussbaumer H., Schmid H.M., Vogel M., 1991, A\&A, 248, 458
\ref M\"{u}rset U., Schmid H.M., 1999, A\&A Supp., 137, 473
\ref Mursula, K, Ulich Th., 1998, Geophysical Research letters, 25, 1837
\ref Ogilvie, G.I., Lesur,  G., 2012, arXiv:1201.5020
\ref Petr\'{i}k K., Hric L., G\'{a}lis R., Friedjung M., Dobrotka A.,  1998, IBVS, 4588, 1
\ref Robinson L.J., 1964, IBVS, 73, 1
\ref Robinson L., 1969, Peremennye Zvezdy, 16, 507
\ref Scargle J.D., 1982, ApJ, 263, 835
\ref Shore S.N. et al., 2010, A\&A, 510, A70
\ref Skopal A., 1998, Contrib. Astron. Obs. Skalnate Pleso, 28, 87
\ref Skopal A., 2001, A\&A, 366, 157
\ref Skopal A., 2007, NewA, 12,597
\ref Skopal A., Chochol D., 1994, IBVS, 4080, 1
\ref Skopal A. et al., 1992, Contrib. Astron. Obs. Skalnate Pleso, 22, 131
\ref Skopal A. et al., 1995, Contrib. Astron. Obs. Skalnate Pleso, 25, 53
\ref Skopal A., Vanko M., Pribulla T., Wolf M., Semkov E., Jones A., 2002, Contrib. Astron. Obs. Skalnate Pleso, 32, 62
\ref Skopal A., Pribulla T., Va\v nko M., Veli\v c Z., Semkov E., Wolf M., Jones A., 2004, Contrib. Astron. Obs. Skalnate Pleso, 34, 45
\ref Skopal A., Va\v{n}ko M., Pribulla T., Chochol D., Semkov E., Wolf M., Jones A., 2007, Astron. Nachrichten, 328, 909
\ref Sokoloski J.L. et al., 2006, ApJ, 636, 1002
\ref Taranova O.G., Yudin B. F., 1982, Azh, 59, 92
\ref van Belle G.T. et al., 1999, AJ, 117, 521
\ref Zahn J.-P., 1977, A\&A, 57, 383
\ref Zamanov R.K., Bode M.F., Melo C.H.F., Bachev R., Gomboc A., Stateva I.K., Porter J.M., Pritchard J., 2007, MNRAS, 380, 1053
\ref Zamanov R. K., 2011, BlgAJ, 15, 19
\ref Zhu Z.X., Friedjung M., Zhao G., Hang H.R., Huang C.C., 1999, A\&A Supp, 140, 69

\label{lastpage}
\end{document}